\documentclass{jsara} \journalinfo{Journal of the Southeastern
  Association for Research in Astronomy, {\bf 6}, 00-00, \today}

\usepackage{pdflscape}
\slugcomment{}

\shorttitle{New Variable Stars in the Cluster NGC 6584}
\shortauthors{Toddy et al.}

\begin{document}

%% This will be set in the editorial office to the starting number of
%% your paper in the journal
\setcounter{page}{01}


\title{New Variable Stars in the Globular Cluster NGC
  6584\altaffilmark{1}}

%% Use \author, \affil, and the \and command to format author and
%% affiliation information.  Note that \email has replaced the old
%% \authoremail command from AASTeX v4.0. You can use \email to mark
%% an email address anywhere in the paper, not just in the front
%% matter.  As in the title, you can use \\ to force line breaks.

\author{Joseph M. Toddy\altaffilmark{2}}
\affil{Department of Physics and Astronomy, University of Georgia,
 Athens, GA 30601}
\and
\author{ Elliott W. Johnson, Andrew N. Darragh, and Brian W. Murphy}
\affil{Department of Physics and Astronomy, Butler University, 
Indianapolis, IN, 46208 \& SARA}

\altaffiltext{1}{Based on observations obtained with the SARA
  Observatory 0.6 m telescope at Cerro Tololo, which is owned and
  operated by the Southeastern Association for Research in Astronomy
  (http://www.saraobservatory.org). }

\altaffiltext{2}{Southeastern Association for Research in Astronomy
  (SARA) NSF-REU Summer Intern} 

\email{bmurphy@butler.edu}

\begin{abstract}
Using the Southeastern Association for Research in Astronomy 0.6 meter
telescope located at Cerro Tololo, we searched for variable stars in
the southern globular cluster NGC 6584. We obtained images during 8
nights between 28 May and 6 July of 2011. After processing the images,
we used the image subtraction package ISIS developed by Alard (2000)
to search for the variable stars. We identified a total of 69 variable
stars in our 10$\times$10 arcmin$^2$ field, including 43 variables cataloged
by Millis \& Liller (1980) and 26 hereto unknown variables. In total,
we classified 46 of the variables as type RRab, with a mean period of
0.56776 days, 15 as type RRc with a mean period of 0.30886 days,
perhaps one lower amplitude type RRe, with a period of 0.26482 days, 4
eclipsing binaries, and 3 long period ($P>2$ days) variable stars.  As
many as 15 of the RRab Lyrae stars exhibited the Bla{\v z}hko
Effect. Furthermore, the mean periods of the RR Lyrae types, the
exhibited period/amplitude relationship, and the ratio of
$N_{c}/(N_{ab}+N_{c})$ of 0.25 are consistent with an Oosterhoff Type
I cluster. Here we present refined periods, V-band light curves, and
classifications for each of the 69 variables, as well as a
color-magnitude diagram of the cluster.
\end{abstract}

%% Keywords should appear after the \end{abstract} command. The
%% uncommented example has been keyed in ApJ style. See the
%% instructions to authors for the journal to which you are submitting
%% your paper to determine what keyword punctuation is appropriate.


%% From the front matter, we move on to the body of the paper In the
%% first two sections, notice the use of the natbib \citep and \citet
%% commands to identify citations.  The citations are tied to the
%% reference list via symbolic KEYs. The KEY corresponds to the KEY in
%% the \bibitem in the reference list below. We have chosen the first
%% three characters of the first author's name plus the last two
%% numeral of the year of publication as our KEY for each reference.

\section{Introduction}

Globular clusters provide an ideal controlled environment to study
variable stars and hence stellar evolution since the stars in the
cluster presumably formed out of the same cloud of gas and dust, have
the same composition, have the same distance, and have the same age
(Smith 2004).  In addition to understanding the important aspects of
late-stage stellar evolution in a controlled environment, variable
stars in globular clusters are one of the primary standard candles
that are used to measure the distances to globular clusters in the
Milky Way and nearby galaxies.  They also provide a method to easily
observe thousands of stars at once, a useful characteristic when
searching for variable stars.

NGC~6584 is a relatively understudied globular cluster located 13.5
Kpc from the Sun in the southern constellation Telescopium (Harris
1996).  It is an Oosterhoff type I cluster (Oosterhoff 1939) and
Sarajedini \& Forrester (1995) have estimated its metallicity at
[Fe/H]=-1.44$\pm$0.16. Early studies of this cluster were performed by
Bailey (1924) who identified nine potential variable stars in its
vicinity.  Millis \& Liller (1980) published the locations of 48
variable stars in NGC~6584. They utilized photographic plates in their
observations and analyses. As a result, the light curves that they
present are relatively incomplete and the periods that they provide
are somewhat imprecise.  Sarajedini \& Forrester (1995) confirmed
these 42 RR Lyrae variables and tentatively identified 56 possible yet
undiscovered RR Lyraes by analyzing the variation in the
frame-to-frame dispersion in V and B-V as a function of V for the
stars in the instability strip of their color-magnitude diagram
(CMD). Those stars which showed a large range in V dispersion were
identified as possible RR Lyraes. In the same year, Samus et
al. (1995), also using photographic photometry, again confirmed Millis
\& Liller's 48 variables.  Since the results of Sarajedini \&
Forrester (1995) and Samus et al. (1995) image subtraction software
and better CCDs have become available.

With the combination of CCD images, an observation window spanning
more than 5 weeks, and an image subtraction method developed by Alard
(2000), we are able to detect and analyze hereto undiscovered variable
stars and produce unprecedented high-quality light curves and periods
for each variable.  In this study we use Alard's image subtraction
method to search the central 10$\times$10 arcmin$^2$ of globular
cluster NGC~6485 for variable stars from observations obtained in May,
June, and July of 2011. We present a preliminary color-magnitude
diagram for the cluster, as well as complete light curves and periods
for 43 of Millis \& Liller's variables and for 26 newly identified
variables.

\section{Observations and Reduction}

We obtained images using the Southeastern Association for Research in
Astronomy (SARA) 0.6 meter telescope at Cerro Tololo Interamerican
Observatory (CTIO). Observations were made on the nights of 28 May; 2,
22, 25, and 28 June; 2, 5, and 6 July 2011 using an Apogee Alta E6
camera with a 1024$\times$1024 pixel Kodak KAF1001E chip, a gain of
1.5 electron per ADU count, and an RMS noise of 8.9 electrons. We used
1$\times$1 binning resulting in an image scale of 0.61{\arcsec}/px
resulting in a 10$\times$10 arcmin$^2$ field of view. The CCD was kept
at a constant temperature of -25 Celsius.  The typical seeing was
1.7{\arcsec} and ranged from 1.2 to 2.8{\arcsec}. On all nights of
observation, 240 second exposures were taken using a Bessel V filter.
The exposure time allowed us to detect dimmer variables but caused 6
of the stars in the field many to be overexposed.  Most of these
overexposed stars appeared to be in the foreground of the cluster.  On
the nights of 2, 5, and 6 July 2011, additional exposures were taken
using Bessel B, R, and I filters with exposure times of 300, 90, and
50 seconds respectively.  All images were debiased, flat-fielded, dark
subtracted, and cleaned of hot, cold, and bad pixels using MaximDL.

\begin{figure}
\centering
\includegraphics[scale=0.47]{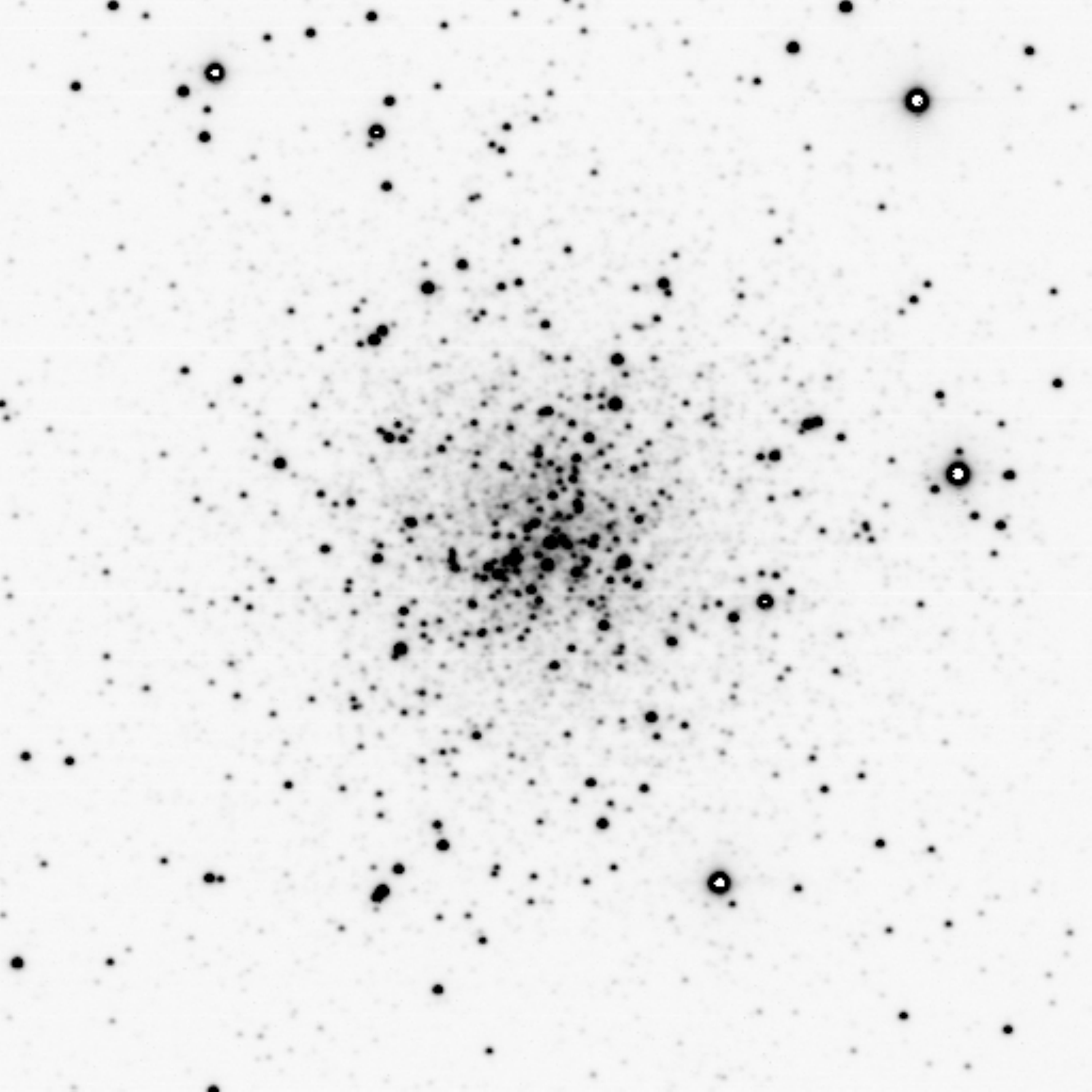}
\caption{The inner 5$\times$5 arcmin$^2$ of the {\tt ref.fits} image
  used for all nights of the analysis/image subtraction.  This image
  is a combination of the best seeing images, with the combined seeing
  near $1.3\arcsec$. Note that the 6 brightest stars in the image are
  overexposed.}
\label{reffits}
\end{figure}

\begin{figure}
\centering
\includegraphics[scale=0.47]{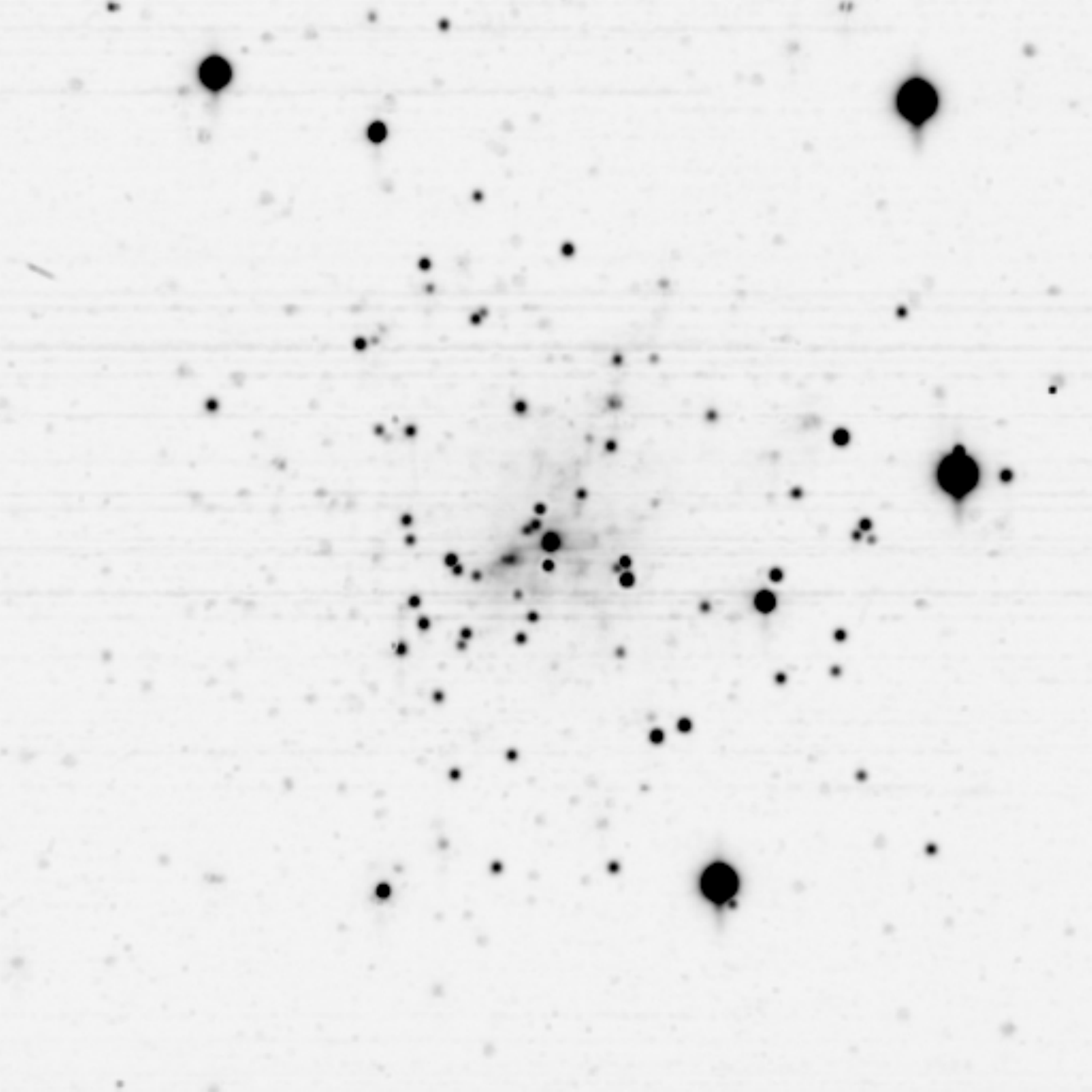}
\caption{The inner 5$\times$5 arcmin$^2$ of the {\tt var.fits} frame
  ouput from ISIS for all eight nights from SARA CTIO.  The relative
  amount of variation is indicated by the brightness of the star.  The
  6 overexposed stars show significant variation only because ISIS is
  unable to fit an accurate point spread function to them. }
\label{varfits}
\end{figure}

\section{Analysis}

Image Subtraction was accomplished using the ISIS-2.2 package (Allard
\& Lupton 1998; Allard 2000). ISIS accounts for changes in seeing
conditions by convolving a high quality reference image to the point
spread functions of each individual frame and then subtracting the two
images to determine any change in intensity.  The package contains
seven c-shell scripts and three parameter files that were used in our
analysis.  The different c-shell scripts are {\tt dates}, {\tt interp}
(registration and interpolation), {\tt ref} (build reference frame),
{\tt subtract} (subtracts images from reference), {\tt detect} (stacks
subtracted images), {\tt find} (finds variables above a user-defined
threshold), and {\tt phot} (produces light curves for all detected
possible variables), along with three parameter files.

We began by running a modified version of the {\tt dates} script,
which identified and registered each image file in the image
directory. We then ran the {\tt interp} script which aligned each
image to a previously chosen alignment image. Next, the {\tt ref}
script combined five of our highest quality images into a reference
frame (Fig. 1). In order to maintain consistent and comparable values
for relative flux across nights, one reference image was created using
five of the best images (those with seeing between 1.2 and
1.3\arcsec). We then ran the {\tt subtract} script, which first
convolves the high-quality reference frame to the point-spread
function of the image to be subtracted (in order to compensate for
variations in seeing over the course of the observations), and then
subtracts the convolved reference frame from each image. These
subtracted images were then stacked using the {\tt detect} script
creating a {\tt var.fits} image which shows the degree of variability
across the observation run for each star in the cluster as shown in
Fig. 2. We then used the {\tt find} script to identify variable stars
above a certain user-defined threshold called {\tt SIGTHRESH}. We used
a {\tt SIGTHRESH} of 0.03 in our analysis to ensure that all possible
variables were found.  This value typically resulted in 1500 possible
variables.  Finally, the {\tt phot} script was used to generate light
curve data files for each identified variable. These light curves
allowed us to eliminate false positives from the {\tt find}
script. Variable stars produce very distinct light curves, and thus
false positives due to noise and variations in seeing were easily
identified amongst these plots and discarded.

Periods were determined, when possible, based on combined photometric
data from all eight nights.  We used the period finding software {\tt
  AVE} (Analisis de Variabilidad Estelar, Analysis of Estellar
Variability) from Grup d'Estudis Astronomics.  {\tt AVE} uses the
Lomb-Scargle algorithm (Scargle 1982) to determine the most probable
period for a given set of data. Due to the more than 5 week
observation span, we were able to determine the periods of most of the
variables with periods less than 1 day to an accuracy of $10^{-5}$
days.  Those variables exhibiting the Bla{\v z}hko Effect were less
accurate due to modulation of their light curves.

Millis \& Liller's (1980) variables were identified visually and
astrometrically and then confirmed by comparing their periods with our
own.  Five of the variables identified by Millis \& Liller were
outside the field of view of our images thus were not be observed.

\begin{figure*}
%counter{figure}{3}
\rotatebox{0}{\includegraphics[scale=0.84]{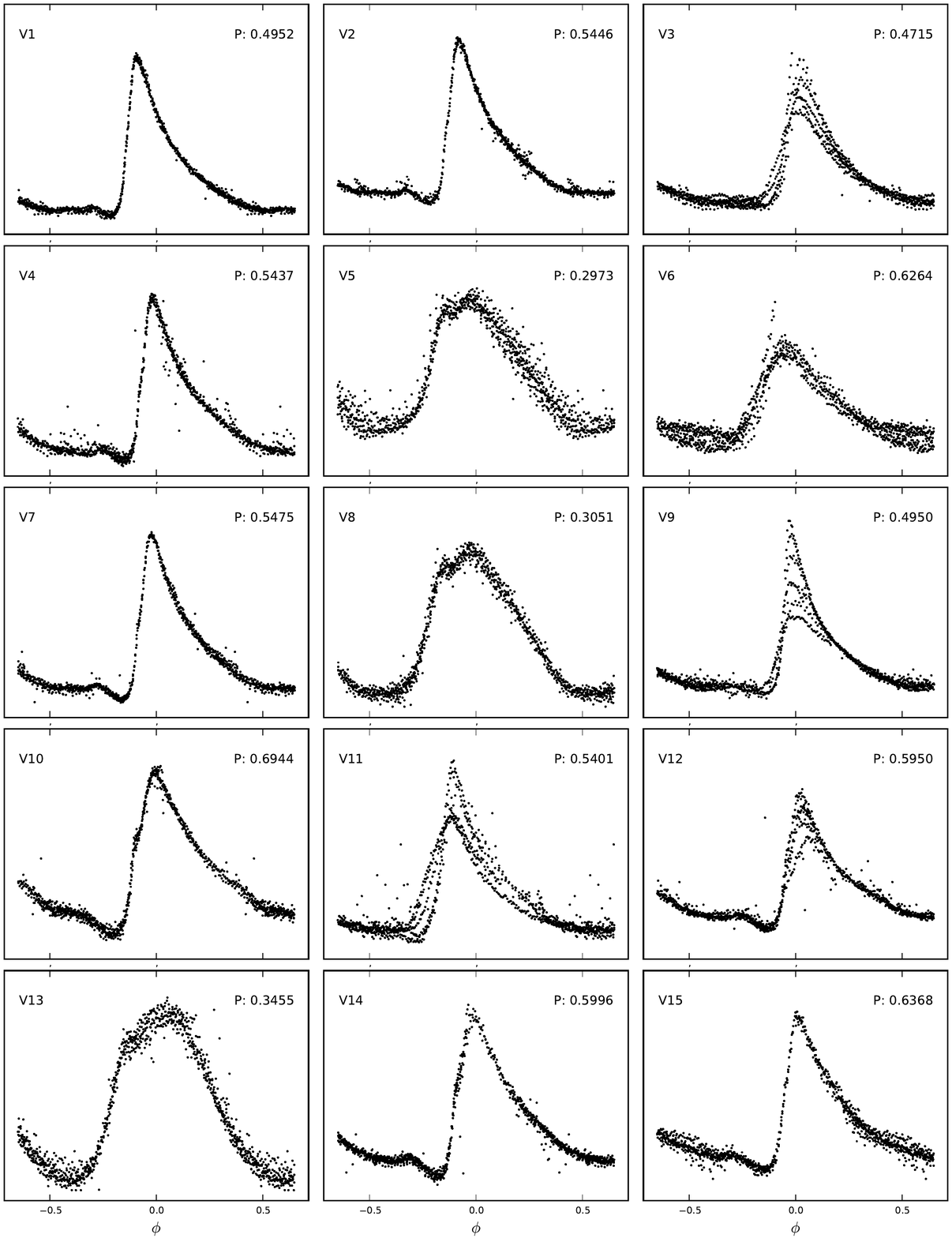}}
\caption{Phased light curves for identifieded variables with periods
  under a day.  The vast majority of these variables are RR Lyrae
  stars.}
\label{phased_variables}
\end{figure*}

\begin{figure*}
\setcounter{figure}{2}
\rotatebox{0}{\includegraphics[scale=0.84]{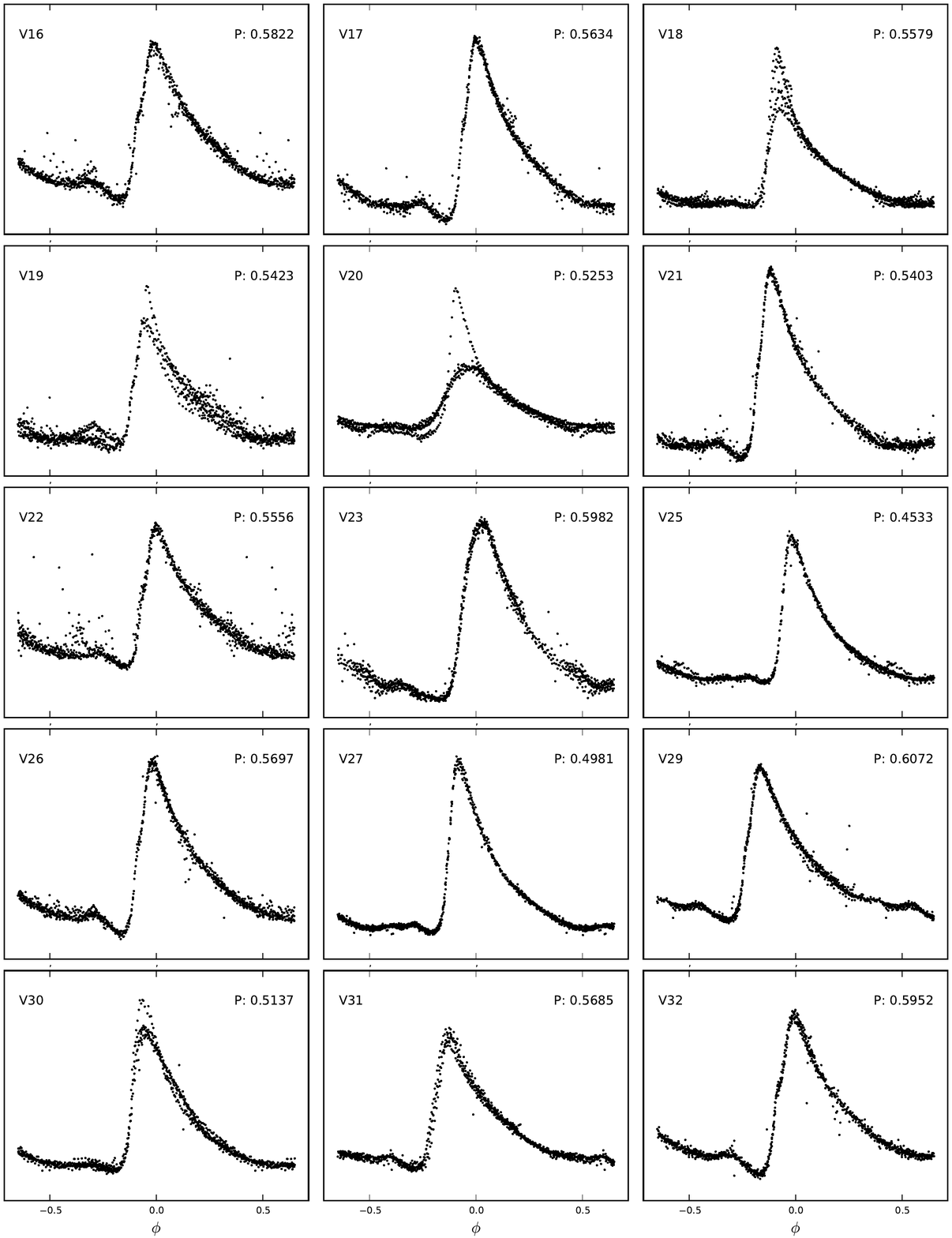}}
\caption{(Continued) Phased light curves for identified variables
  with periods under a day.  The vast majority of these variables are
  RR Lyrae stars.}
\label{phased_variables}
\end{figure*}

\begin{figure*}
\setcounter{figure}{2}
\rotatebox{0}{\includegraphics[scale=0.84]{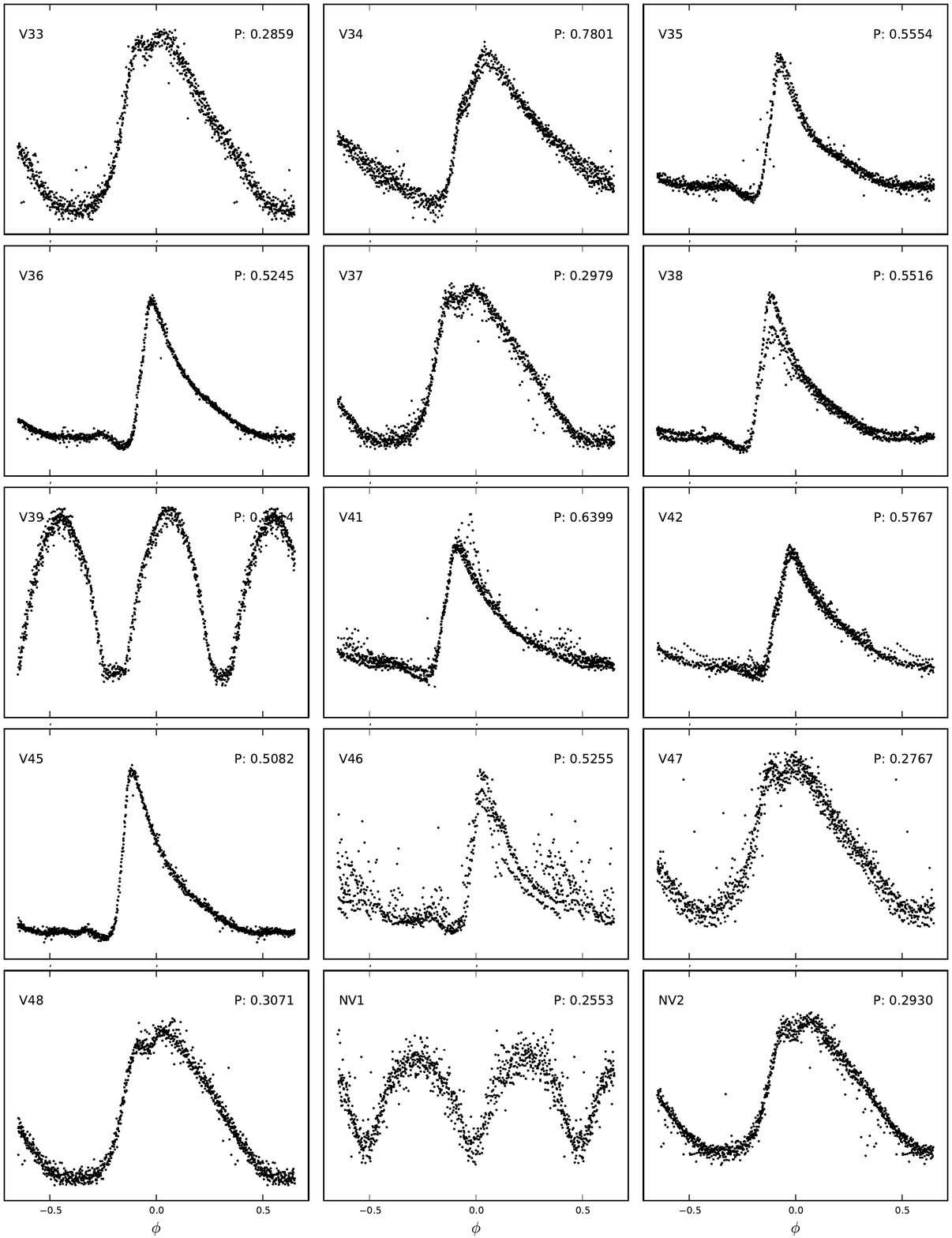}}
\caption{(Continued) Phased light curves for identified variables
  with periods under a day.  The vast majority of these variables are
  RR Lyrae stars.}
\label{phased_variables}
\end{figure*}

\begin{figure*}
\setcounter{figure}{2}
\rotatebox{0}{\includegraphics[scale=0.84]{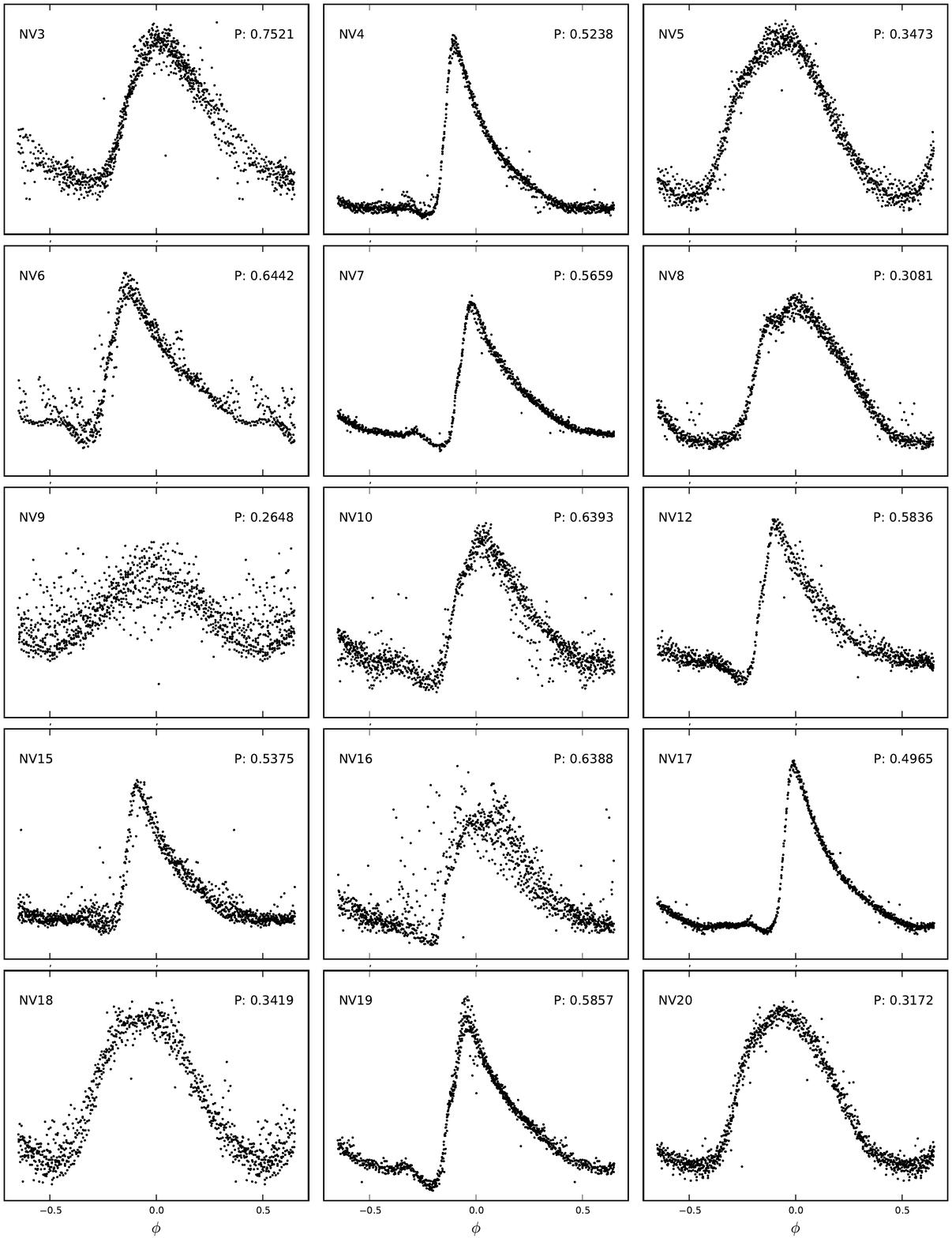}}
\caption{(Continued) Phased light curves for identified variables
  with periods under a day.  The vast majority of these variables are
  RR Lyrae stars.}
\label{phased_variables}
\end{figure*}

\begin{figure*}
\centering
\setcounter{figure}{2}
{\includegraphics[scale=0.84]{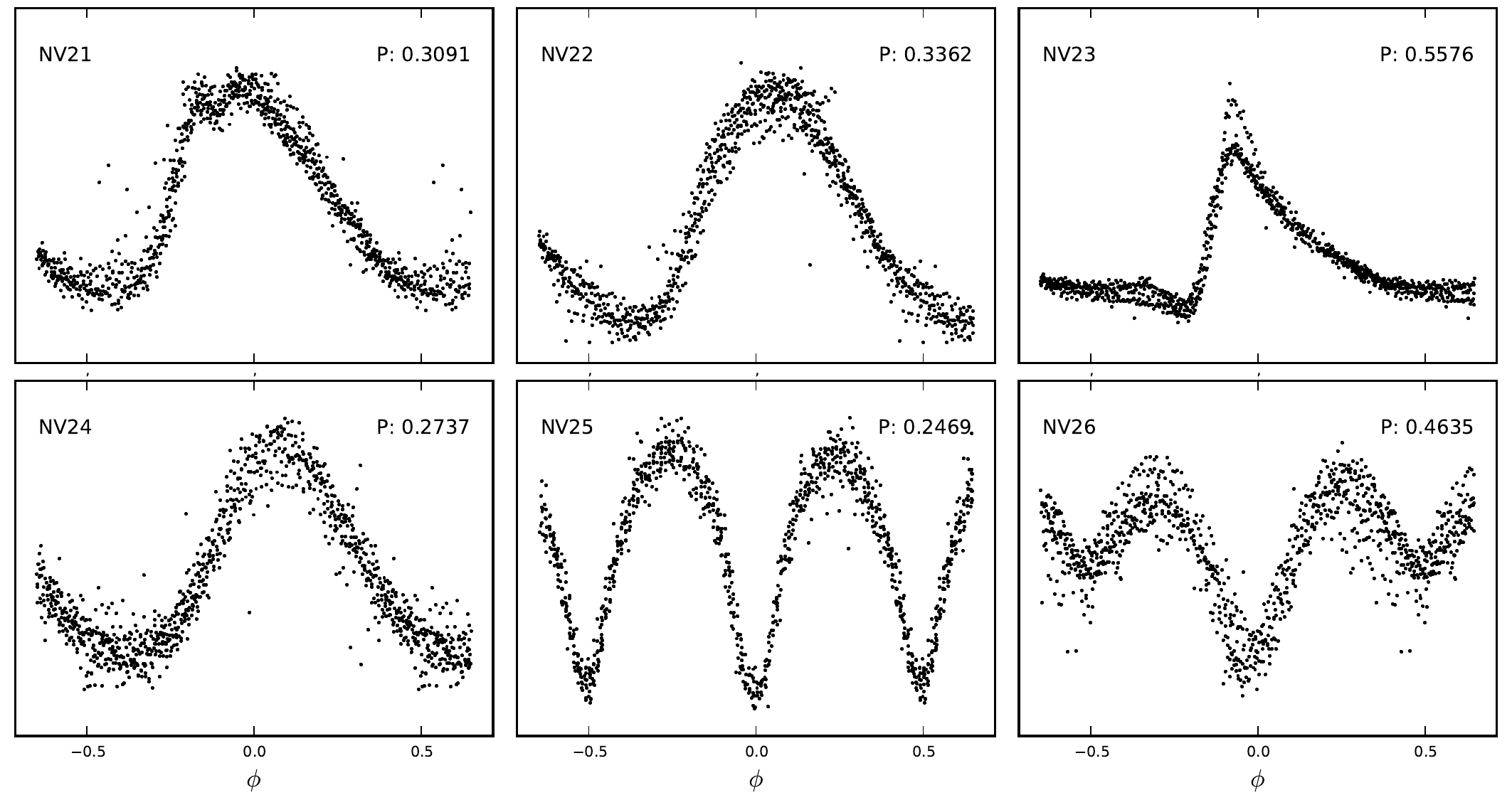}}
\caption{(Continued) Phased light curves for identified variables
  with periods under a day.  The vast majority of these variables are
  RR Lyrae stars.}
\label{phased_variables}
\end{figure*}

\begin{figure*}
\centering
{\includegraphics[scale=0.79]{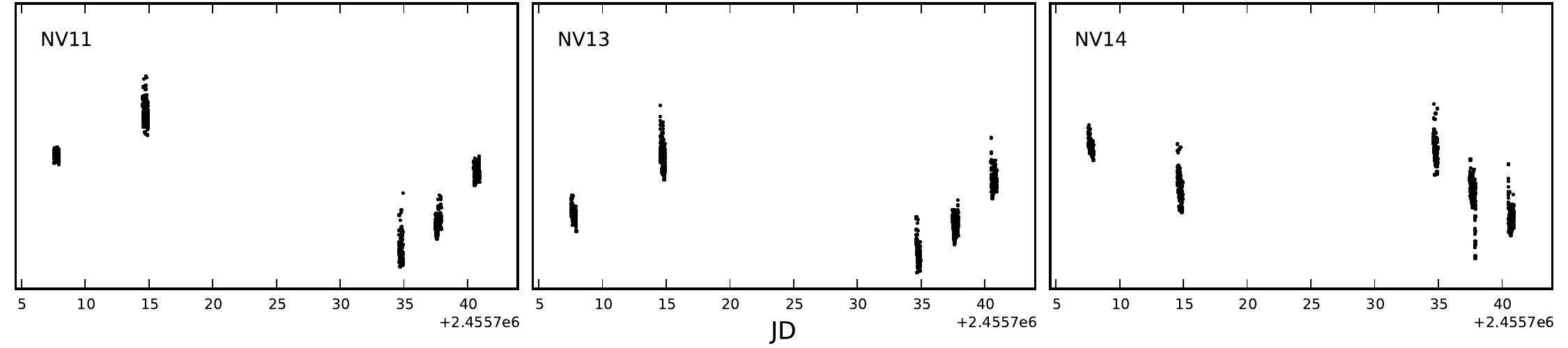}}
\caption{Light curves for the 3 long period variables found in our
  analysis.}
\label{lpvs}
\end{figure*}

\begin{figure*}
\centering
\includegraphics[scale=0.83]{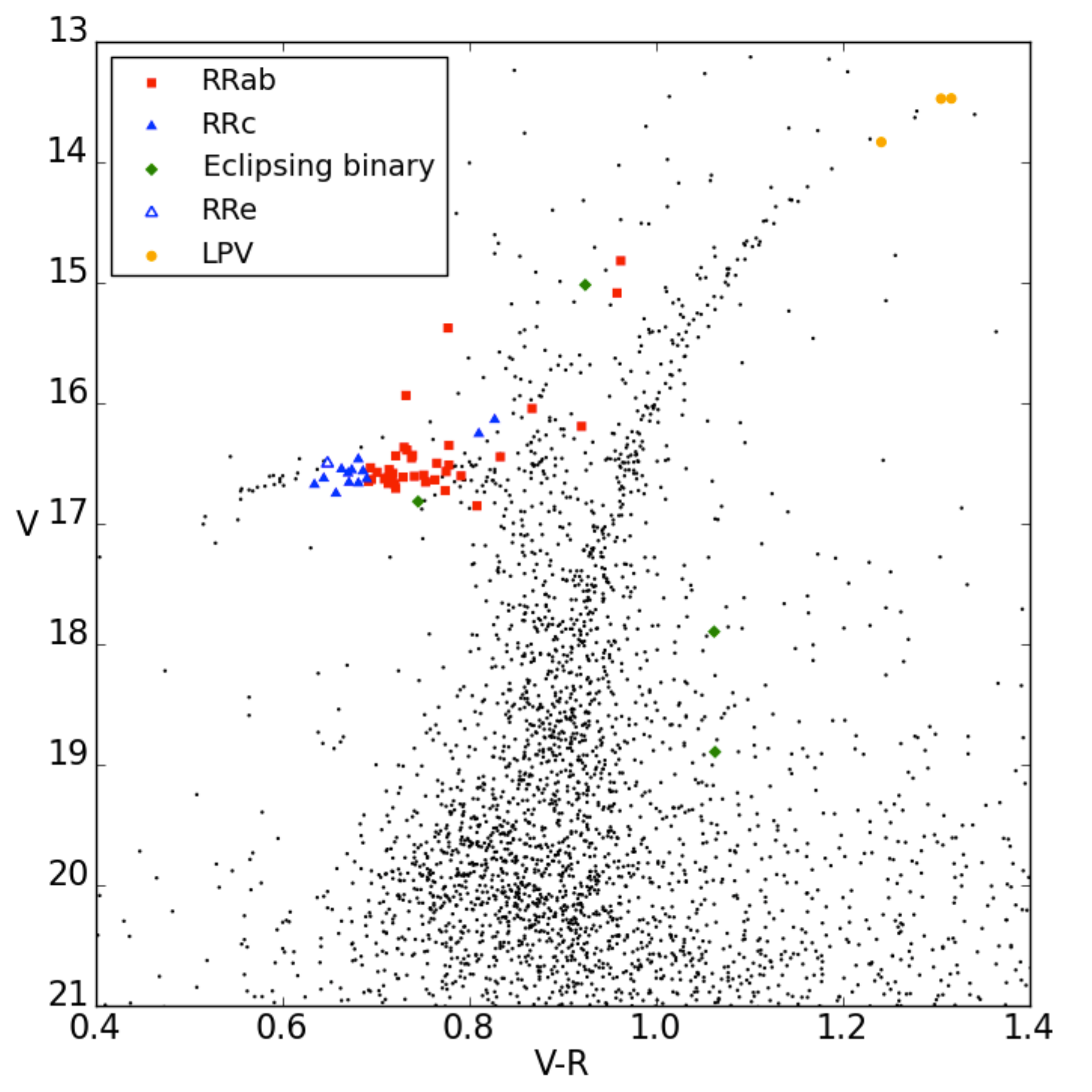}
\caption{Color-magnitude diagram of NGC 6584, with variable stars
  indicated.  Note that the RRc (blue) variables are bluer than the
  RRab's (red). All but 6 of the variables found appear to be cluster
  members. The two eclipsing variables to the right of the giant
  branch of the cluster are likely not cluster members.  The 2 RRc and
  4 RRab Lyrae variables just to the right of the instability strip
  are all blends.  The 4 brightest RRab's may be foreground objects.}
\label{period_hist}
\end{figure*}

\section{Results}

\begin{deluxetable}{lrc}
%\tabletypesize{\scriptsize}
\tablecaption{Classification Summary}
\tablewidth{0pc}
\tablehead{
\colhead{Variable Type} & \colhead{Count} & \colhead{Period (days)} 
}
\startdata
RRab&46&0.56776\\
RRc&15&0.30886\\
RRe&1&0.26482\\
Eclipsing&4&0.32431\\
LPV&3& $>20$
\enddata
\end{deluxetable}

Table 1 summarizes our findings by classification.  Table 2 list the
positions, periods, and classification of all of the variables that we
identified.  In Table 2 variables previously identified by Millis \&
Liller (1980) are labled with a V and the number they assigned in the
first column.  Newly discovered variables are labeled as NV and are in
order of increasing right ascension. We detected all the Millis \&
Liller (1980) variables that were within our 10$\times$10 arcmin$^2$
field.  Their variables 24, 27, 40, 43, and 44 were outside our field
of view. We did note that variable 45 was misidentified on the finder
chart in Millis \& Liller (1980), thus the position we list for of
this variable has been shifted by several arc-seconds in Table 2.  We
found periods for variables V12, V19, and 39 that Millis \& Liller
were unable to determine.  We also were able to eliminate those
periods where multiple possible periods were listed; specifically V5,
V20 and V35.  Besides the increased accuracy of our periods the only
significant change from Millis and Liller was V13 which we found to be
0.34549 days as compared to 0.39 days.

Each detected variable was classified based on the shape of its light
curve, its period, and its maximum amplitude.  In addition to the
Miller \& Liller variables we found 26 additional variables in our
field, most of which are RR Lyrae variables, long period variables, or
eclipsing variables.  Of the total 69 variables detected, 46 were
classified as RRab stars, 15 as RRc, 1 possible RRe, 4 eclipsing
binaries, and 3 long period variables.  From the fraction
$N_c/(N_{ab}+N_c)$ we confirm that the cluster is of Oosterhoff Type
I.  RRab stars are characterized by large amplitudes and periods
between 0.4 and 0.8 days. RRc stars have smaller amplitudes and
periods between 0.2 and 0.4.  NV9 has the characteristics of a type
RRe variable (Alcock et al. 1996; Bono et al. 1997).  It has a lower
amplitude and period than type RRc stars.  Eclipsing binary stars are
easily identified by their unique light curves. As many as 15 of the
RRab stars appeared to exhibit the so-called Bla{\v z}hko effect
(Bla{\v z}hko 1907; Smith 2004), which is characterized by a long term
modulation in amplitude and period. These variables are noted with and
asterisk in Table 2. Their periods are likely to be less accurate
given their behaviour.  We have not assigned periods to the long
period variables at the tip of the asymptotic giant branch because
further observations will be needed to establish the their periods. We
assign 'LPV' to these variables with a period greater than 2 days.

\renewcommand{\thefootnote}{\arabic{footnote}}

\setlongtables
\begin{longtable}{rcccccrcc}
\tabletypesize{\scriptsize}
\tablecaption{Variables Stars in NGC 6584}
\tablewidth{0pt}
\tablehead{
\colhead{V$\#$} & \multicolumn{3}{c}{RA (h,m,s)} & 
\multicolumn{3}{c}{Dec
 ($^{\circ}$,$\arcmin$,$\arcsec$)} &
\colhead{Period (d)} & \colhead{Type}
}
\startdata
V1   &  18&  18& 28.82& -52&  13& 27.0&  0.49520 &   RRab\\
V2   &  18&  18& 42.20& -52&  12& 30.0&  0.54465 &   RRab\\
{\tablenotemark{a}V3}
     &  18&  18& 42.14& -52&  13& 16.2&  0.47162 &   RRab\\
V4   &  18&  18& 40.30& -52&  13& 25.5&  0.54377 &   RRab\\
{\tablenotemark{b}V5}
     &  18&  18& 40.49& -52&  13& 29.1&  0.29725 &   RRc\\
{\tablenotemark{a}V6}   
     &  18&  18& 34.45& -52&  12&  7.8&  0.62619 &   RRab\\
V7   &  18&  19&  1.34& -52&  11& 55.3&  0.54748 &   RRab\\
V8   &  18&  18& 55.62& -52&  13& 35.3&  0.30511 &   RRc\\
{\tablenotemark{a}V9}
     &  18&  18& 43.55& -52&  12&  2.7&  0.49513 &   RRab\\
V10  &  18&  18& 37.07& -52&  11& 36.7&  0.69432 &   RRab\\
{\tablenotemark{a}V11}  
     &  18&  18& 35.42& -52&  13& 10.8&  0.53945 &   RRab\\
{\tablenotemark{a}V12} 
     &  18&  18& 35.84& -52&  12& 32.6&  0.59505 &   RRab\\
V13  &  18&  18& 35.85& -52&  14& 32.3&  0.34549 &   RRc\\
{\tablenotemark{a}V14}
     &  18&  18& 41.20& -52&  13& 43.2&  0.59954 &   RRab\\
V15  &  18&  18& 23.56& -52&  12& 41.9&  0.63632 &   RRab\\
V16  &  18&  18& 42.04& -52&  12& 58.6&  0.58220 &   RRab\\
V17  &  18&  12& 41.87& -52&  13& 22.3&  0.56338 &   RRab\\
{\tablenotemark{a}V18}
     &  18&  18& 32.99& -52&  13& 18.3&  0.55789 &   RRab\\
{\tablenotemark{a}V19}
     &  18&  18& 40.66& -52&  13&  6.4&  0.54223 &   RRab\\
{\tablenotemark{a,b}V20} %b 
     &  18&  18& 28.15& -52&  12& 57.9&  0.52530 &   RRab\\
V21  &  18&  18& 34.48& -52&  13& 55.1&  0.54025 &   RRab\\
V22  &  18&  18& 38.02& -52&  12& 50.5&  0.55559 &   RRab\\
V23  &  18&  18& 30.76& -52&  13& 10.0&  0.59812 &   RRab\\
V25  &  18&  18& 56.91& -52&  11& 12.8&  0.45336 &   RRab\\
V26  &  18&  18& 33.62& -52&  13& 52.1&  0.56970 &   RRab\\
V28  &  18&  18& 48.09& -52&  12& 19.8&  0.49815 &   RRab\\
V29  &  18&  18& 41.49& -52&  11& 40.5&  0.60720 &   RRab\\
{\tablenotemark{a}V30}
     &  18&  18& 42.85& -52&   9& 31.9&  0.51375 &   RRab\\
V31  &  18&  18& 30.61& -52&  13& 38.5&  0.56831 &   RRab\\
V32  &  18&  18& 40.71& -52&  14&  5.4&  0.59515 &   RRab\\
V33  &  18&  18& 26.81& -52&  11& 55.2&  0.28595 &   RRc\\
V34  &  18&  18& 30.11& -52&  12& 16.4&  0.78010 &   RRab\\
{\tablenotemark{b}V35}  
     &  18&  18& 36.82& -52&  12& 48.4&  0.55550 &   RRab\\
V36  &  18&  18& 38.94& -52&  14&  0.2&  0.52454 &   RRab\\
V37  &  18&  18& 26.06& -52&  14& 28.1&  0.29786 &   RRc\\
{\tablenotemark{a}V38}
     &  18&  18& 26.81& -52&  11& 55.1&  0.55159 &   RRab\\
V39  &  18&  18& 19.48& -52&  10& 47.2&  0.33141 &   EB\\
{\tablenotemark{a}V41}  
     &  18&  18& 40.70& -52&  13&  4.9&  0.63992 &   RRab\\
V42  &  18&  18& 39.88& -52&  11& 55.3&  0.57675 &   RRab\\
{\tablenotemark{c}V45}
     &  18&  18& 28.75& -52&  12& 30.8&  0.50823 &   RRab\\
V46  &  18&  18& 25.08& -52&  12& 35.3&  0.52555 &   RRab\\
V47  &  18&  18& 38.32& -52&  13& 20.8&  0.27666 &   RRc\\
V48  &  18&  18& 38.55& -52&  13& 26.1&  0.30706 &   RRc\\
NV1  &  18&  18&  6.83& -52&  12& 25.0&  0.25531 &   EB\\
NV2  &  18&  18& 19.26& -52&  12& 38.3&  0.29297 &   RRc\\
NV3  &  18&  18& 27.84& -52&  12& 59.4&  0.75210 &   RRab\\
NV4  &  18&  18& 27.99& -52&  12& 55.7&  0.52384 &   RRab\\
NV5  &  18&  18& 28.19& -52&  14&  7.0&  0.34724 &   RRc\\
NV6  &  18&  18& 32.16& -52&  14& 43.2&  0.64406 &   RRab\\
NV7  &  18&  18& 32.79& -52&  12& 24.7&  0.56583 &   RRab\\
NV8  &  18&  18& 33.14& -52&  10& 15.5&  0.30807 &   RRc\\
NV9  &  18&  18& 34.47& -52&  12& 49.1&  0.26482 &   RRe\\
NV10 &  18&  18& 35.57& -52&  13& 31.4&  0.63930 &   RRab\\
NV11 &  18&  18& 35.66& -52&  12& 20.6&   $>20$  &   LPV\\
NV12 &  18&  18& 35.67& -52&  13&  7.1&  0.58387 &   RRab\\
NV13 &  18&  18& 37.69& -52&  12& 59.8&   $>20$  &   LPV\\
NV14 &  18&  18& 37.79& -52&  13&  6.2&   $>20$  &   LPV\\
{\tablenotemark{a}NV15} 
     &  18&  18& 38.16& -52&  12& 54.8&  0.53741 &   RRab\\
NV16 &  18&  18& 38.42& -52&  12& 56.0&  0.63630 &   RRab\\
NV17 &  18&  18& 38.58& -52&  12& 21.4&  0.49644 &   RRab\\
NV18 &  18&  18& 38.77& -52&  13& 14.1&  0.34180 &   RRc\\
{\tablenotemark{a}NV19} 
     &  18&  18& 39.48& -52&  14& 32.8&  0.58572 &   RRab\\
NV20 &  18&  18& 39.93& -52&  11& 21.3&  0.31722 &   RRc\\
NV21 &  18&  18& 40.01& -52&  13&  9.1&  0.30911 &   RRc\\
NV22 &  18&  18& 40.81& -52&  10& 32.5&  0.33625 &   RRc\\
{\tablenotemark{a}NV23} 
     &  18&  18& 41.98& -52&  12& 53.7&  0.55739 &   RRab\\
NV24 &  18&  18& 42.72& -52&  12& 28.9&  0.27374 &   RRc\\
NV25 &  18&  18& 44.98& -52&   8& 10.7&  0.24689 &   EB\\
NV26 &  18&  18& 51.42& -52&  13& 30.8&  0.46364 &   EB

\end{longtable}
\footnotetext[a]{variable exhibitiing Bla{\v z}kho Effect.}
\footnotetext[b]{variables with significantly revised periods.}
\footnotetext[c]{misidentified by Millis \& Liller (1980).}\enddata

Cluster membership was determined using a color magnitude diagram
(CMD). The CMD was constructed from 137 V and 155 R frames from
several nights to determine magnitudes, colors, and cluster membership
of the variables.  {\tt DAOPHOT} was used to determine V and R
magnitudes of the stars (Stetson 1987).  In the CMD shown in Figure 5
RRab and RRc variables are indicate by the red and blue colored
points, respectively.  RRc variables tend to lie on the warmer (left)
side of the instability strip whereas RRab variables are to the cooler
(right) of the instability strip. Scatter in the variables is partly
due to blending of the variables with other stars, our averaging of
several hundred images before using {\tt DAOPHOT}, and possible
variable reddening in the direction of the cluster. 

Six of our variables, V11, V22, NV7, NV19, NV20, and NV21 were blended
with other stars and show significant offset to the right of the
instability strip and are likely cluster members.  Four of the RRab
variables V15, V41, NV4, and NV15 were significantly brighter than the
RR Lyrae variables found in the instability strip of the cluster and
are suspected not to be cluster members. From the CMD we are able to
determine that 2 of the eclipsing variables to the right of the giant
branch are not cluster members. The 3 long period variables that we
found all lie near the tip of the asymptotic giant branch.  The
remainder of the RR Lyrae stars lie very near or in the instability
strip and thus are likely cluster members.

\section{Conclusions}

Over a 39 day time span from late May to early July of 2011 we
observed the globular cluster NGC 6584 for 8 nights using the SARA 0.6
meter telescope at CTIO. Using the image subtraction software ISIS we
were able to find precise periods of previously known variables and 26
newly discovered variables stars, most of which are RR Lyrae stars.
Within our 10$\times$10 arcmin$^2$ field of view we detected all the
variables listed by Millis \& Liller (1980) bringing the total to 74
variables in the vicinity of the cluster.  Of all the RR Lyrae
variables all but 4 appeared to be members of the cluster.  Two of the
4 eclipsing binaries are likely to be cluster members.  In total, we
classified 46 of the variables as type RRab, with a mean period of
0.56776 days, 15 as type RRc with a mean period of 0.30886 days,
perhaps one lower amplitude type RRe, with a period of 0.26482 days, 4
eclipsing binaries, with a mean period of 0.3243 days, and 3 long
period ($P>2$ days) variable stars.  As many as 15 of the RRab Lyrae
stars exhibited the Bla{\v z}hko Effect. The ratio of
$N_c/(N_{ab}+N_c)$ of 0.25 is consistent with cluster being of type
Oosterhoff Type I.

%\end{multicols}

%% separate it off from the body of the text using the
%% \acknowledgments command.

%% Included in this acknowledgments section are examples of the AASTeX
%% hypertext markup commands. Use \url without the optional [HREF]
%% argument when you want to print the url directly in the
%% text. Otherwise, use either \url or \anchor, with the HREF as the
%% first argument and the text to be printed in the second.

\acknowledgments

We thank C.  Alard for making ISIS 2.2 publically available.  This
project was funded in part by the National Science Foundation Research
Experiences for Undergraduates (REU) program through grant NSF
AST-1004872. Additionally A. Darragh, E. Johnson, B.  Murphy were
partially funded by the Butler Insitute for Research and Scholarship.
The authors also thank F.  Levinson for a generous gift enabling
Butler University's membership in the SARA consortium.

% More information on the AASTeX macros package are
% available at \url{http://ucpjournals.uchicago.edu/AAS/AASTeX/}.

%% The reference list follows the main body and any appendices.  Use
%% LaTeX's thebibliography environment to mark up your reference list.
%% Note \begin{thebibliography} is followed by an empty set of curly
%% braces.  If you forget this, LaTeX will generate the error "Perhaps
%% a missing \item?".  Note - get these in the right form by going to
%% ADS, searching for the papers you are citing, and after the
%% abstract find ``Preferred format for Abstract'' thebibliography
%% produces citations in the text using \bibitem-\cite
%% cross-referencing. Each reference is preceded by a \bibitem command
%% that defines in curly braces the KEY that corresponds to the KEY in
%% the \cite commands (see the first section above).  Make sure that
%% you provide a unique KEY for every \bibitem or else the paper will
%% not LaTeX. The square brackets should contain the citation text
%% that LaTeX will insert in place of the \cite commands.

%% We have used macros to produce journal name abbreviations.  AASTeX
%% provides a number of these for the more frequently-cited journals.

%% delimit the year from the author name used in the citation.  See
%% the natbib documentation for more details and options.

\end{document}